# Pressure induced metallization of $Cu_3N$

Wen Yu[a], Linyan Li and Changqing Jin[b]

Institute of Physics, Chinese Academy of Sciences,

PO Box 603, Beijing 100080, People's Republic of China

We employed accurate full-potential density-functional theory and linearized augmented plane wave (FPDFT-LAPW) method to investigate the electronic properties and possible phase transitions of $Cu_3N$ under high pressure. The anti-perovskite structure $Cu_3N$ is a semiconductor with a small indirect band gap at ambient pressure. The band gap becomes narrower with increasing pressure, and the semi-conducting anti-$ReO_3$ structure undergoes a semiconductor – semimetal transition at pressure around 8.0 GPa. At higher pressure, a subsequent semimetal – metal transition could take place above 15GPa with a structural transition from anti-$ReO_3$ to $Cu_3Au$ structure.

[a] Also at: Physics Department, University of Science and Technology Beijing, Beijing 100080, P. R. China
Electronic mail: wenyu@sas.ustb.edu.cn

[b] Correspondence author:

Electronic mail: CQJIN@aphy.iphy.ac.cn

Motivated by their applications in electronic-optical industry, interest has been grown in fabricating copper nitride ($Cu_3N$) films. [1,2] The crystal structure of $Cu_3N$ is of the anti-$ReO_3$ type ( a perovskite missing A site atom) with a simple cubic unit cell of lattice constant 3.817Å as shown in Fig.1. In this structure, the copper atoms occupy the center of the cubic edges forming collinear bonds with two nearest neighbor anions instead of occupying the face-centered cubic close-packing sites. As a consequence, this crystal structure has many vacant interstitial sites. This open crystal structure is suited for the interposition of other elements or compression under high-pressure conditions.

Electronic structure calculation has shown that pure $Cu_3N$ is semiconductor with a small indirect band gap of about 0.23eV while $Cu_3N$ with Pd interposition exhibits semi-metallic behavior.[3] Recent theoretical calculation also confirmed the metallic property of $Cu_3N$ with extra Cu interposition.[4] The effect of lattice parameters on the electronic properties of $Cu_3N$ were also discussed in reference 4 with an emphasis on the increase of the energy gap with the increase of the lattice parameters.

Compared with bulk $ReO_3$, of which the high-pressure properties have been studied extensively both experimentally and theoretically, [5-7] little work has been done on bulk $Cu_3N$, especially under high pressure conditions. In order to understand the electronic nature of this material under high pressure, we performed first principles calculations on $Cu_3N$ with both ideal anti-$ReO_3$ type (denoted type I hereafter) and a hypothetic $Cu_3Au$ type (denoted type II hereafter) structures. The space group is the same for the two structures (Pm3m), but the Cu and N atoms occupy different wyckoff sites. For the anti-$ReO_3$ structure, N is in 1a sites and Cu in 3d sites; for the $Cu_3Au$ structure, N is still in 1a sites, but Cu in 3c sites. The $Cu_3Au$ is expected to be favored under high-pressure condition because of the close-packed nature of this structure.

We employed accurate full-potential density-functional theory (DFT) and the full potential linearized augmented plane wave (FP-LAPW) method as implemented in WIEN2K code to investigate the electronic properties and possible phase transition of $Cu_3N$ under high pressure.[8] The exchange and correlation effects were treated using the Generalized Gradient Approximation (GGA) of Perdew et. al..[9] The calculated value using the GGA approximation is in general closer to the experiments than the one calculated using LDA. Spin orbit coupling is neglected. 6000 $k$ points were used for the Brillouin-zone integrations of the two structures.

The static structural properties such as lattice constant, total energy, and bulk modulus can be obtained from the calculated total energy as a function of volume. We have calculated total energies of type I and II $Cu_3N$ structures for volumes ranging from 0.58 to $1.10O_{exp}$, where $O_{exp}$ is the experimental equilibrium volume of type I structure (lattice parameter 3.817Å), they are then least-squares-fitted to the Murnaghan equation of state.[10] The minimum total energy, the equilibrium lattice constant, and the bulk modulus are readily deduced from the fitted parameters in the equation of state. The calculated lattice constants, total energies, bulk moduli and their pressure derivatives are listed in Table I where they are compared with available experiments and other *ab initio* calculations. The fitted total-energy curves as a function of volume for the two phases are shown in Fig. 2. It is well known that the phase transformation occurs when the Gibbs free energy becomes equal between the two phases. By applying this theorem to the zero-temperature case considered here, it is easily shown that the pressure induced phase transformation occurs along the common tangent line between the energy curves of the two phases under consideration and the negative of the slope of the common tangent line is the transition pressure. The calculated transition pressure and volumes are also given in Table I. Since the transition pressure is only 17GPa, it is possible to identify the high-pressure phase with a diamond anvil cell experiment.

In order to understand the electronic nature of $Cu_3N$, we have calculated the band structures and density of states (DOS) for type I and type II structures. The energy bands with Cu and N characters and total and partial DOS for type II structure are displayed in Figure 3. It is obvious from Figure 3 that type II structure is metallic. The energy bands can be divided into four parts: a low lying Cu $3d$ – N $2p$ bonding part with energy less than –3.8eV; an intermediate Cu $3d$ non-bonding part of about 2.8eV width; an upper Cu $3d$ – N $2p$ anti-bonding part with energies ranging from –1.0eV to 1.8eV and higher conduction bands. The energy bands crossing Fermi level consist mainly of contribution from the anti-bonding Cu $3d$ – N $2p$ states.

The energy bands for type I structure at equilibrium and semiconductor – semimetal transition volumes are shown in Figure 4. The indirect semi-conducting band gap at equilibrium is about 0.27eV which is in good agreement with other theoretical calculations.[3,4] The lowest energy of the conduction bands is at M point and the highest energy of the valence bands is at R point. The band gap decreases with volume. At transition point from type I to type II structures, a conduction band crosses the Fermi level at M point and overlaps with the valence band at R point, leading to a semi-metallic, rather than metallic behavior of the system. The onset pressure of this behavior where the conduction band is coincident with the fermi level is about 8 GPa. The total and partial DOS for type II structure at equilibrium and transition volumes are given in Figure 5. It can be found that at transition point, there is no energy gap at Fermi level and although very small, the DOS is non-zero.

From the above calculation and discussion, we conclude for the first time that, under high pressure, electronic phase transition for anti-$ReO_3$ structure $Cu_3N$ from semi-conductor to semi-metal occurs and even higher pressure will favor a metal state accompanying a structural transition from anti-$ReO_3$ structure to $Cu_3Au$ structure. This could be useful to tune the physical properties of $Cu_3N$ through external applied pressure or just the strain force in the thin film.

Acknowledgement:

This work was supported by NSFC (grant 50332020, 50328102, 50321101), and the Ministry of Science and Technology of China (2002CB613301).

FIGURE CATIONS:

FIGURE 1. The schematic view of the anti perovskite type crystal structure of $Cu_3N$ highlighting the Cu6N octahedron.

FIGURE 2. Total Energy - volume relations for anti-$ReO_3$ and $Cu_3Au$ structures of $Cu_3N$ (Energy and volume with respect to the anti-$ReO_3$ structure at the measured equilibrium volume).

FIGURE 3. Band structure with characters for (a) Cu and (b) N and total and partial density of states (c) for $Cu_3Au$ structure.

FIGURE 4. Band structure with characters for (a) Cu and (b) N at equilibrium and (c) Cu and (d) N at 90% measured volumes for Anti-$ReO_3$ structure.

FIGURE 5. Density of states for anti-$ReO_3$ structure at (a) equilibrium and (b) 90% measured volumes.

TABLES:

TABLE I. Calculated $Cu_3N$ equilibrium structural properties and transition pressure, volumes from anti-$ReO_3$ to $Cu_3Au$ structure. Volumes are normalized to the measured zero-pressure volume of anti-$ReO_3$ structure.

| Structure | Anti-$ReO_3$ | $Cu_3Au$ |
|---|---|---|
| Lattice constant $a_0$ (Å) | 3.826(3.82[a])(3.817[b]) | 3.50 |
| Bulk modulus $B_0$ (GPa) | 116(104[a]) | 153 |
| Pressure derivative $B_0'$ | 4.47(5.26[a]) | 4.74 |
| Total energy $E_0$ (eV)[c] | 0.0 | 1.33 |
| Transition Pressure (GPa) | 17 | |
| Transition Volume | 0.905 | 0.703 |

[a]Reference 4. [b]Reference 3. [c]Total energy per formula unit with respect to the calculated equilibrium energy of anti-$ReO_3$ structure.

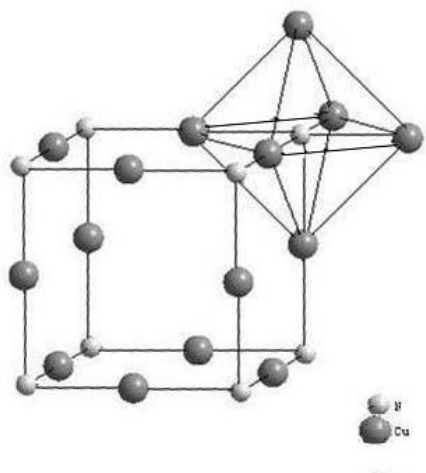

Figure 1

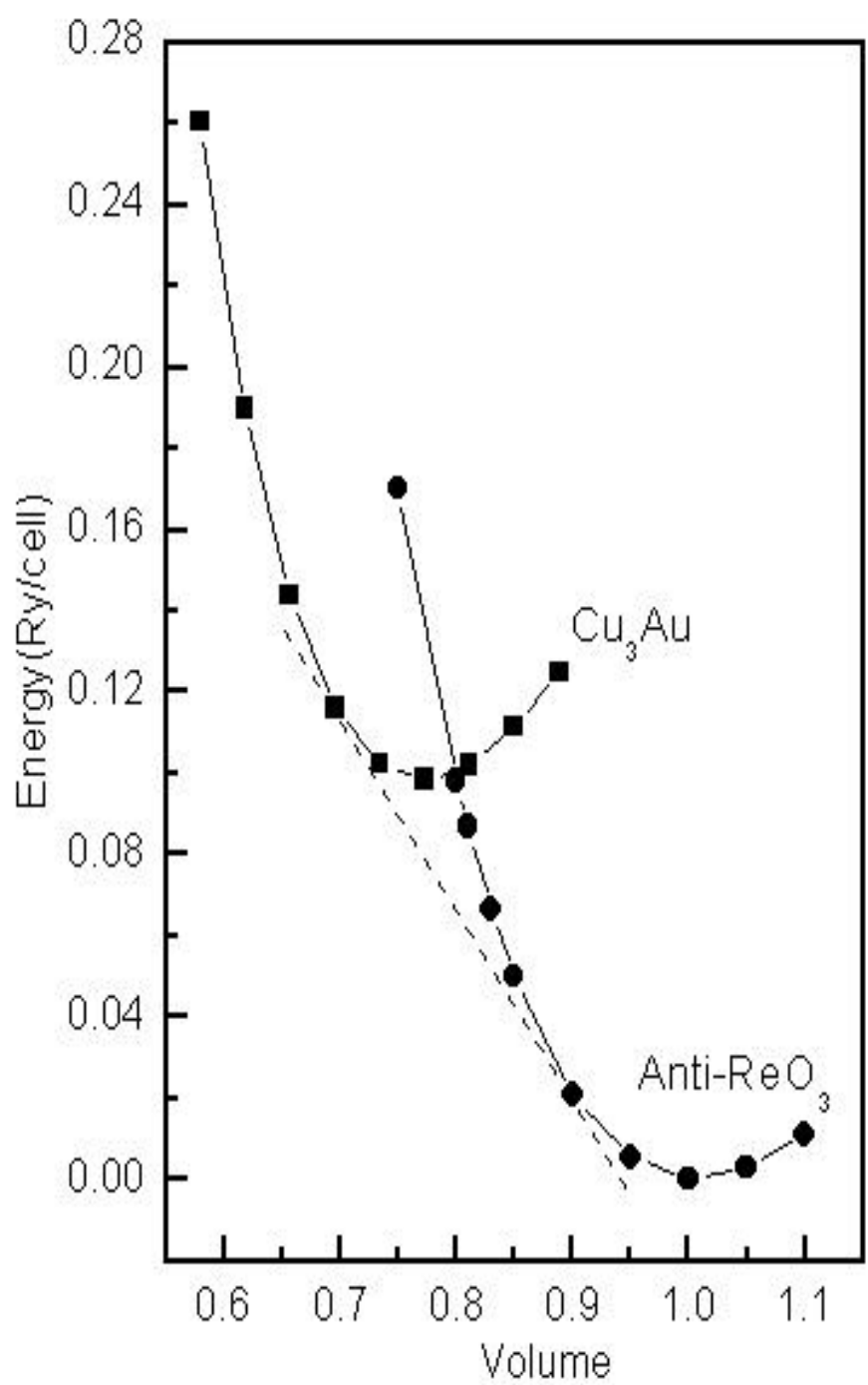

Figure 2

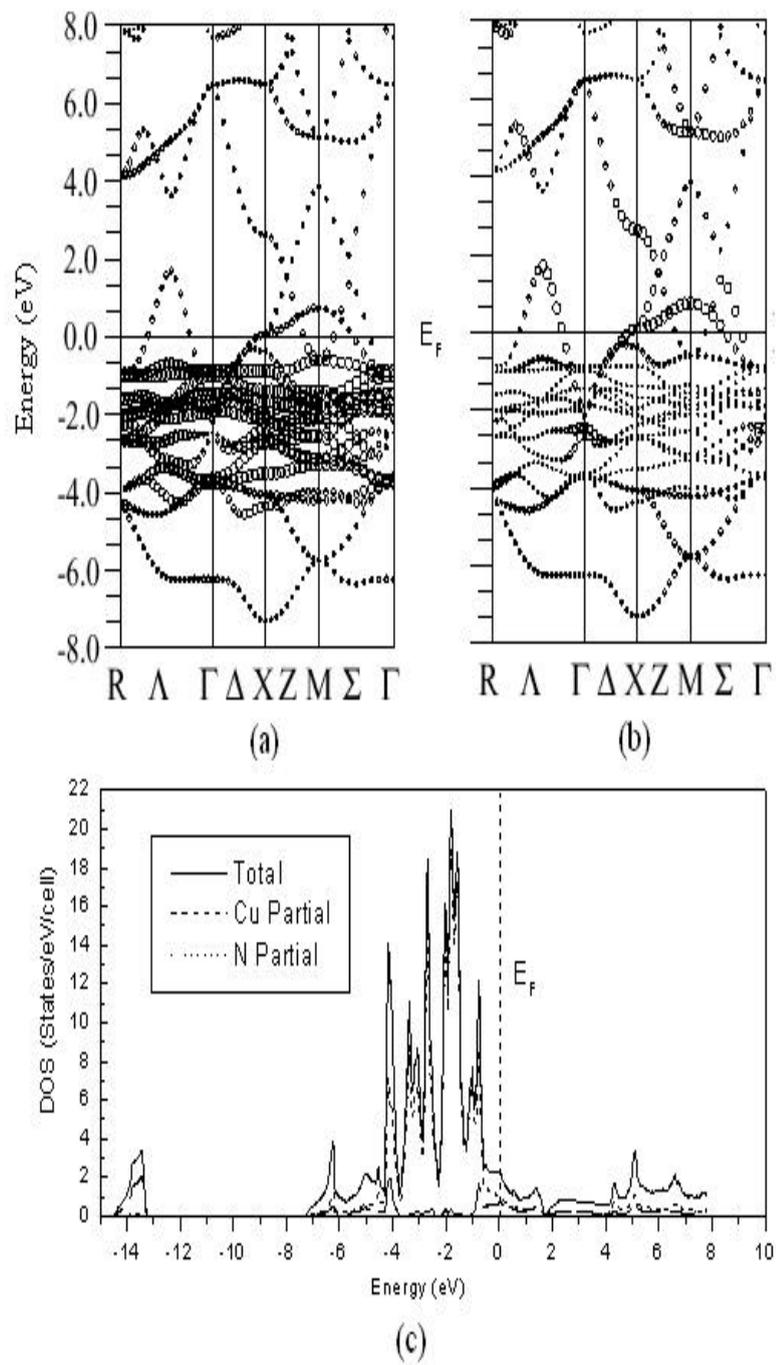

Figure 3

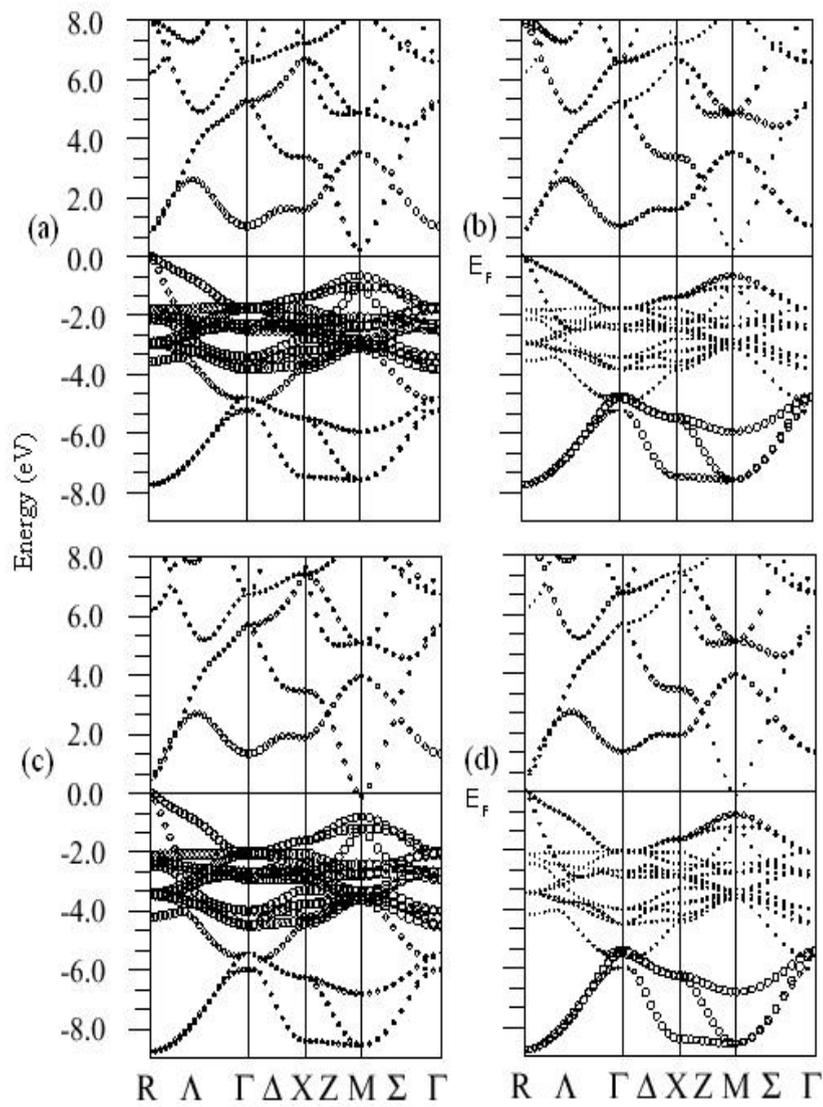

Figure 4

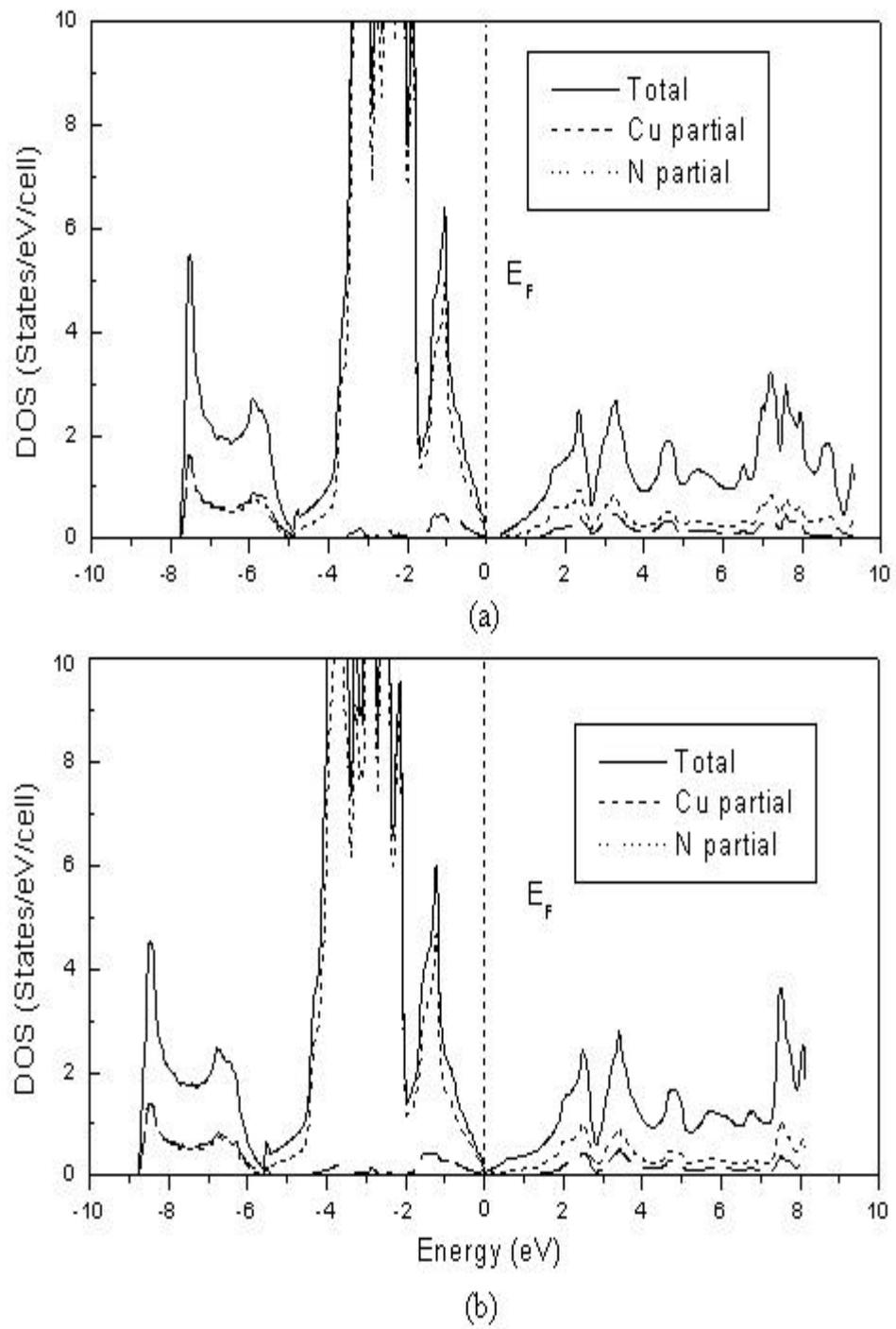

Figure 5